# Special flow model for passive particle transport considering internal noise

Boris S. Maryshev[1,2], Lyudmila S. Klimenko[1,3]

[1]Institute of Continuous Media Mechanics UB RAS

[2]Nizhny Novgorod State University, Nizhny Novgorod, 603022, Russia

[3]Perm State University, Perm, Russia

E-mail: bmaryshev@mail.ru, lyudmilaklimenko@gmail.com

We have generalized the semi-analytic approach of special flow to the description of flows of passive particles taking into account internal noise. The model is represented by a series of recurrence relations. The recurrence relations are constructed by numerically solving the Langevin equations in the presence of a random force, for an ensemble of passive particles during transport through a secluded cell. This approach allows us to estimate the transit time dependence near stagnation points for fluid elements carried by the flow. Such estimates are obtained for the most important types of stagnation points. It is shown that macroscopic transport of an ensemble of particles through such a lattice is possible only when internal noise is taken into account. For Gaussian and non-Gaussian noise at low intensity the transit time has one peak, which is a consequence of the existence of vortices of one stagnation point. Increase of noise intensity leads to slowing down of particle transport.

## 1. Introduction

The transport process in a fluid flow is usually described by the Diffusion-Advection equations(ADE) [1]. These two processes can be separated from each other in the presence of macroscopic flow. However, if the flow has a complex structure (for example, [2-3]), then the averaged characteristics are used to describe the flow. In this case, all additional transport processes arising due to the flow microstructure are also referred to as diffusion, i.e. turbulent diffusion [4] or effective diffusivity [5] for transport through a porous medium. Papers [6, 7] estimate the flow contribution to the diffusion process based on various assumptions about the flow microstructure. From these papers, one key condition on the flow structure can be emphasized: distributions of flow velocities should be such that for the transport of particles in this flow the first momentum (ensemble averaged over time) and the second momentum (ensemble averaged over space) exist and have finite values. For specific systems, this condition is not always true, and as a result, the various deviations from the standard diffusion law are observed.

In the present work, we have generalized the well-known [9] analytical solution of the system of equations describing two-dimensional flow of a viscous fluid in the presence of an additional force. For this purpose, we model complex flows by systems of several vortices. To begin with, we restrict ourselves to time-independent two-dimensional flows of incompressible viscous fluid generated in regions of simple geometry by periodic forces. The flow through some stagnation points is described in terms of a special flow [8]. Due to the incompressibility of the fluid, flow lines can be related to trajectories for an integrable Hamiltonian system. The movement of a fluid element along the flow line from the neighborhood of one stagnation point to another can be represented as a mapping, since the position in the neighborhood of any stagnation point is completely determined by the initial position. The time it takes the fluid element to move from the initial position to the current position can be defined as the sum of the travel times of all stagnant points in the vicinity of the flow line to the current position. The travel time of a fluid element in the vicinity of a stagnation point depends on the type of stagnation point and is estimated for most typical points.

The novelty of the present study is also related to the fact that the problem is solved taking into account the internal Gaussian and non-Gaussian (Cauchy distribution) noise.

## 2. The flow for two dimensional laminar flow of viscous fluid with additional force

The plane stationary flow of the viscous incompressible fluid obeys the Navier-Stokes equations

$$u\frac{\partial u}{\partial x}+w\frac{\partial u}{\partial y}=-\frac{1}{\rho}\frac{\partial p}{\partial x}+\nu\frac{\partial^2 u}{\partial x^2}+\nu\frac{\partial^2 u}{\partial y^2}+F_x$$
$$u\frac{\partial w}{\partial x}+w\frac{\partial w}{\partial y}=-\frac{1}{\rho}\frac{\partial p}{\partial y}+\nu\frac{\partial^2 w}{\partial x^2}+\nu\frac{\partial^2 w}{\partial y^2}+F_y, \quad (1)$$
$$\frac{\partial u}{\partial x}+\frac{\partial w}{\partial y}=0$$

where $(x, y)$ are the coordinates in the plane, $(u, w)$ is a vector of fluid velocity, $p(x, y)$ is pressure, the vector $(F_x, F_y)$ denotes density of external force, and, finally, $\rho$ and $\nu$ are, respectively, density and kinematic viscosity of the fluid. It is convenient to rewrite the problem using the stream function $\psi(x, y)$, introduced by $u=-\partial\psi/\partial y$ and $w=\partial\psi/\partial x$. In terms of $\psi$ the problem (1) becomes

$$\frac{\partial\psi}{\partial x}\frac{\partial\Delta\psi}{\partial y}-\frac{\partial\psi}{\partial y}\frac{\partial\Delta\psi}{\partial x}=\nu\Delta\Delta\psi+\frac{\partial F_y}{\partial x}-\frac{\partial F_x}{\partial y}$$
$$\Delta=\frac{\partial^2}{\partial x^2}+\frac{\partial^2}{\partial y^2} \quad . \quad (2)$$

In a stationary flow, the fluid particles or passive tracers move along nonintersecting streamlines (isolines of the stream function). We consider a square cell of the size $H\times H$. If the particle of passive tracer at the time moment $t=0$ is placed in the point $x=x_0$, $y=0$ then this particle will reach the coordinate $y=H$ in point $x=X\ x_0$ in the time moment $t=T\ x_0$. It means that process of such particle transport through the cell can be determined by the functions $X\ x_0$ and $T\ x_0$ which are determined by the force $(F_x, F_y)$. Due to this fact we can introduce the special flow as the mapping of outlet coordinate on inlet coordinate $x_i = X\ x_{i-1}$ with the passage time function $t_i = T\ x_{i-1}$. Due to the periodical boundary conditions $u,w|_{x=0}=u,w|_{x=H}$ and $u,w|_{y=0}=u,w|_{y=H}$ it is naturally to consider the periodical force $(F_x, F_y)$ as following

$$\int_0^H F_x dy = 0, \quad \int_0^H F_y dx = 0. \quad (3)$$

For the periodical force the equation (1) has the solutions in the form $u,w = \alpha/H, \beta/H$, where $\alpha$ and $\beta$ are arbitrary constants which have the meaning as the intensities of mean drift along $x$ and $y$ directions respectively. In terms of mean drift intensities, we can partially write the solution for the stream function in the form

$$\psi = \frac{\beta}{H}x - \frac{\alpha}{H}y + \phi(x,y), \quad \int_0^h \phi(x,y)\,dxdy = 0., \qquad (4)$$

The periodical boundary conditions with the solution (4) allow us to consider the mapping $x_i = X(x_{i-1})$ as the simple shift on the 2-torus in form of $x_i = (x_{i-1} + \rho) \mod 1$, where the $\rho = \alpha/\beta$ is a mean displacement along $x$ axis at one iteration. As a result the special flow is a mapping on the 2-torus with an additional passage time function $t_i = T(x_{i-1})$.

### 2.1 Nonlinear flow without stagnation points

By applying the some elaborate periodic forcing, it is possible to excite a velocity field with a single vortex [9]; in term of $\phi$, a typical stream function reads

$$\phi(x,y) = a\sin(kx)\cos(ky) - a\sin(kx) - a\cos(ky), \qquad (5)$$

Where $k = 2\pi/H$ is the period of solution and $a$ is the amplitude of spatial flow modulation, after shifting the coordinates $x \to x/H + 0.75$, $y \to y/H + 0.5$ we obtain

$$\phi(x,y) = a\sin(2\pi(x+0.75))\cos(2\pi(y+0.5)) - a\sin(2\pi(x+0.75)) - a\cos(k(y+0.5)), \qquad (6)$$

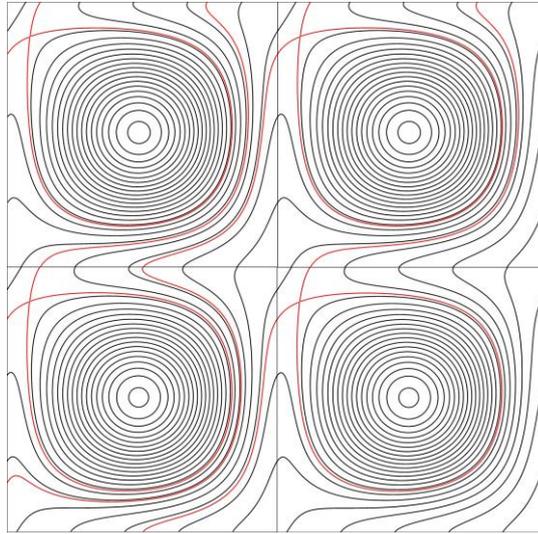

Fig.1. Example of velocity field obeys Eq.(6), with $a=1$

Hence, the equations

$$\frac{\partial x}{\partial t} = u(x,y) = \alpha - \frac{\partial \phi}{\partial y}; \quad \frac{\partial y}{\partial t} = w(x,y) = \beta + \frac{\partial \phi}{\partial x}, \qquad (7)$$

define the trajectory of fluid element or passive tracer in the flow. The condition of zero velocity $u = w = 0$ defines the special trajectory – the stagnation point. The solution of equations (7) with zero flow velocity exists only when the value of amplitude $a$ is larger than critical value $a_{cr}$ which is given by

$$a_{cr} = \frac{\sqrt{-6(\alpha^2+\beta^2)(\alpha^2+6\alpha\beta+\beta^2)(\alpha^2-6\alpha\beta+\beta^2)+6\sqrt{(\alpha^4+14\alpha^2\beta^2+\beta^4)^3}}}{72\pi\alpha\beta}. \quad (8)$$

For the values $a < a_{cr}$ the stagnation points in the considered flow are absent. The function $\phi(x,y)$ (6) defines the periodic flow pattern with spatial period 2 along $x$ and $y$ axes. The isolines of such flow for the case $a < a_{cr}$ are shown in Fig. 2.

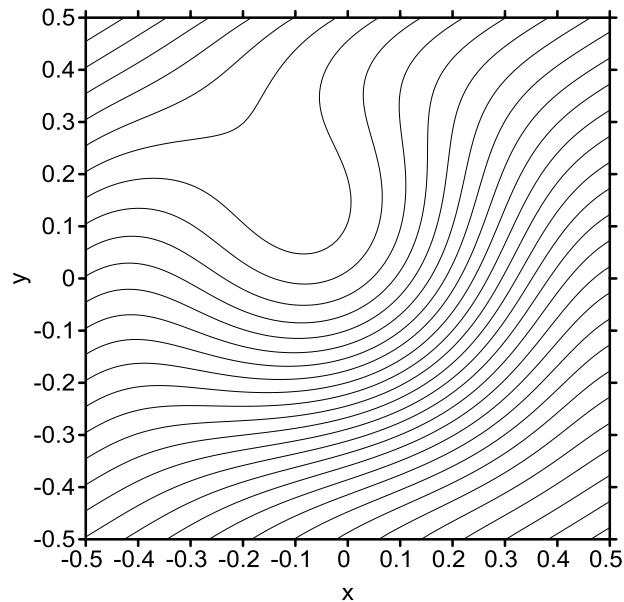

Fig. 2 Flow without stagnation points at $\alpha = 1$, $\beta = 0.5\sqrt{5} - 0.5$, and amplitude $a = 0.1 < a_{cr} \approx 1.005$.

The flow configuration shown in Fig. 2 does not contain stagnation points. However, the latter contains regions of slow motion, which leads to dispersion of fluid elements or passive tracer in the fluid.

**2.2 Isolated stagnation points**

Let us consider the flow in form (6) when the force amplitude $a = a_{cr}$. In this case the solution of equation (7) gives the coordinates of the stagnation points in the form

$$\begin{aligned}\alpha &= 2\pi a_c \sin(2\pi(y_s+0.5))(\sin(2\pi(x_s+0.75))-1) \\ \beta &= 2\pi a_c \cos(2\pi(x_s+0.75))(\cos(2\pi(y_s+0.5))-1)\end{aligned}, \quad (9)$$

The isolines of corresponding flow into the one spatial period $-0.5 \leq x \leq 0.5$ and $-0.5 \leq y \leq 0.5$ and for $a = a_{cr}$ are presented in Fig.3 .

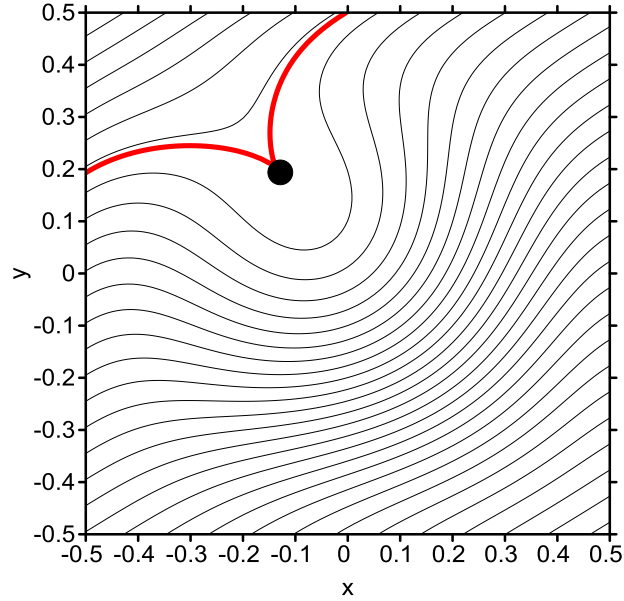

Fig. 3 Flow with stagnation points at $\alpha = 1$, $\beta = 0.5\sqrt{5} - 0.5$, and amplitude $a = a_{cr} \approx 0.100518$. Streamlines of the flow (6) Red curves indicate the trajectories with infinite passage time

Let us consider more regular case when $a > a_{cr}$ the solution for the stagnation point is given by the equations

$$\begin{aligned}\alpha &= 2\pi a \sin\left(2\pi\left(y_s + 0.5\right)\right) \sin(2\pi\left(x_s + 0.75\right) - 1) \\ \beta &= 2\pi a \cos\left(2\pi\left(x_s + 0.75\right)\right) \cos(2\pi\left(y_s + 0.5\right) - 1)\end{aligned}, \quad (12)$$

The flow map into the one spatial period for $a = 1 > a_{cr}$ is shown in Fig.4

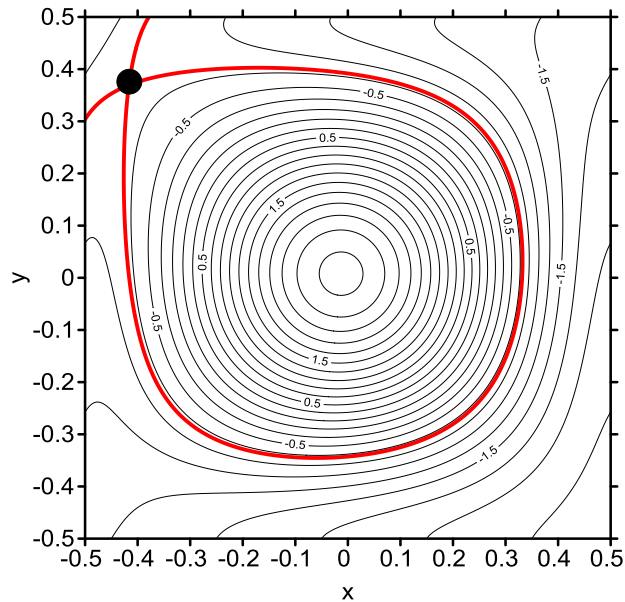

Fig. 4 Flow with stagnation points at $\alpha = 1$, $\beta = 0.5\sqrt{5} - 0.5$, $\nu = 1$ and force amplitude $a = 1 > a_{cr}$. Streamlines of the flow (6) Red curves indicate the trajectories with infinite passage time.

## 3. Particle transport equations

The motion of a passive particle with initial position in $x^0, y^0$ and suspended in the flow (5) can be described by the following Langevin equations

$$\begin{aligned} x^{i+1} &= x^i + u(x^i, y^i)\tau + \sqrt{2D\tau}G^i, \\ y^{i+1} &= y^i + w(x^i, y^i)\tau + \sqrt{2D\tau}Q^i, \\ t^i &= i\tau, \end{aligned} \quad (13)$$

where $\tau$ is the time step, $i$ is the step number, $D$ is the diffusion coefficient, $G^i, Q^i$ are the independent random variables with Gaussian distribution. Since the domain under study is covered with a lattice consisting of identical cells of the period $l$, it makes sense to use the approach developed in [5]. In the absence of diffusion, when a particle passes through one cell in a particular direction, its position at the exit of the cell is uniquely described by its initial position. Let us fix the initial coordinate $x$ for any particle, so let $x_0 = x^0 = -0.5$. We consider the passage from $x_0$ to $x_1 = 0.5$. In this case, the coordinate at the exit $y_1 = H(y_0 = y^0)$ as well as the passage time for one cell $t_1 = T(y_0 = y^0)$ are uniquely determined by the flow (2). Thus, the coordinates and passage time for $n$ cells can be determined as a sum of the corresponding mappings

$$x_n = n - 0.5, \quad y_N = y_0 + \sum_{j=0}^{n-1} H(y_j), \quad t_N = \sum_{j=0}^{n-1} T(t_j). \quad (14)$$

This approach is called the construction of a special flow. The single-valued functions $H$ and $T$ exist only in the case when there is no random factor associated with the diffusion. However, if we consider transport through a sufficiently long chain of elementary cells, then it is possible to use a special flow in the form (5), only functions $H$ and $T$ should be replaced by their expected values. To obtain the expected values we have calculated the particle motion with floating $y_0 = y^0$, and fixed $x_0 = x^0 = -0.5$ up to the time moment $t = t^i$, when the particle reaches $x^i = 0.5$. Averaging over the value of the coordinate and time over the set of realizations, we obtain

$$\begin{aligned} \langle T \rangle &= T(y_0) = \frac{1}{M}\sum_{j=1}^{M} t^{i,j}, \quad \langle y \rangle = H(y_0) = y_1 = \frac{1}{M}\sum_{j=1}^{M} y^{i,j}, \\ x^{i,j} &= 0.5, \quad x^{0,j} = -0.5, \quad y^{0,j} = y^0, \end{aligned} \quad (15)$$

where $j$ is the realization number, $M$ is the full number of random realizations.

The obtained functions $T(y_0)$ and $y_1 = H(y_0)$ are presented in Fig.5. It is seen that at small diffusivity $D$ the dependence $T(y_0)$ has one peak which is the sequence of existence of one stagnation point vortices (see fig. 4). The increase of diffusion leads to homogenization of dependences. The effect of transport slowing down for trajectories near stagnation points decreases.

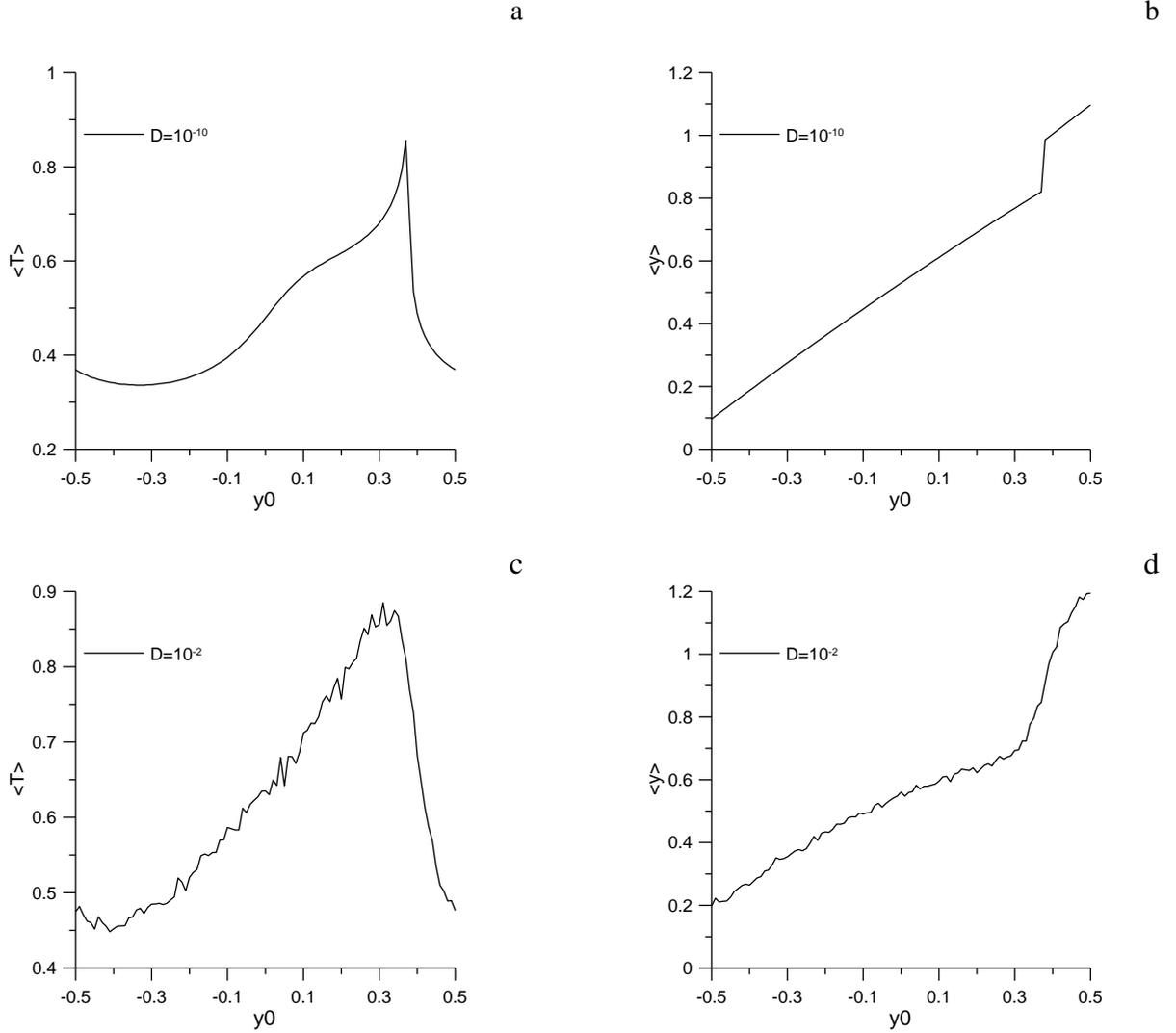

Fig 5 The dependence of the average time and coordinate on $y_0$ for different diffusivity of Gaussian noise

The mapping in form (5) contains the deterministic functions $T(y_0)$ and $y_1 = H(y_0)$, but the description of diffusive process means that we should add the random displacement and rewrite the mapping (5) in the form

$$y_j = y_{j-1} + H(y_{j-1}) + \sqrt{D_y(y_{j-1})}\, G_j,$$
$$t_j = T(y_j) + \sqrt{D_T(y_{j-1})}\, S_j \qquad (1)$$
$$x_n = n - 0.5, \quad y_n = \sum_{j=0}^{n-1} y_j, \quad t_N = \sum_{j=0}^{n-1} t_j.$$

The $D_y(y_0)$ and $D_T(y_{j-1})$ are the dispersions of coordinate and time and it should be also calculated by

$$D_t(y_0) = \frac{1}{M}\sum_{j=1}^{M}(t^{i,j} - T(y_0))^2,$$
$$D_y(y_0) = \frac{1}{M}\sum_{j=1}^{M}(y^{i,j} - H(y_0))^2. \quad (2)$$

The examples of such functions for small and great values of diffusivity are presented in Fig.6

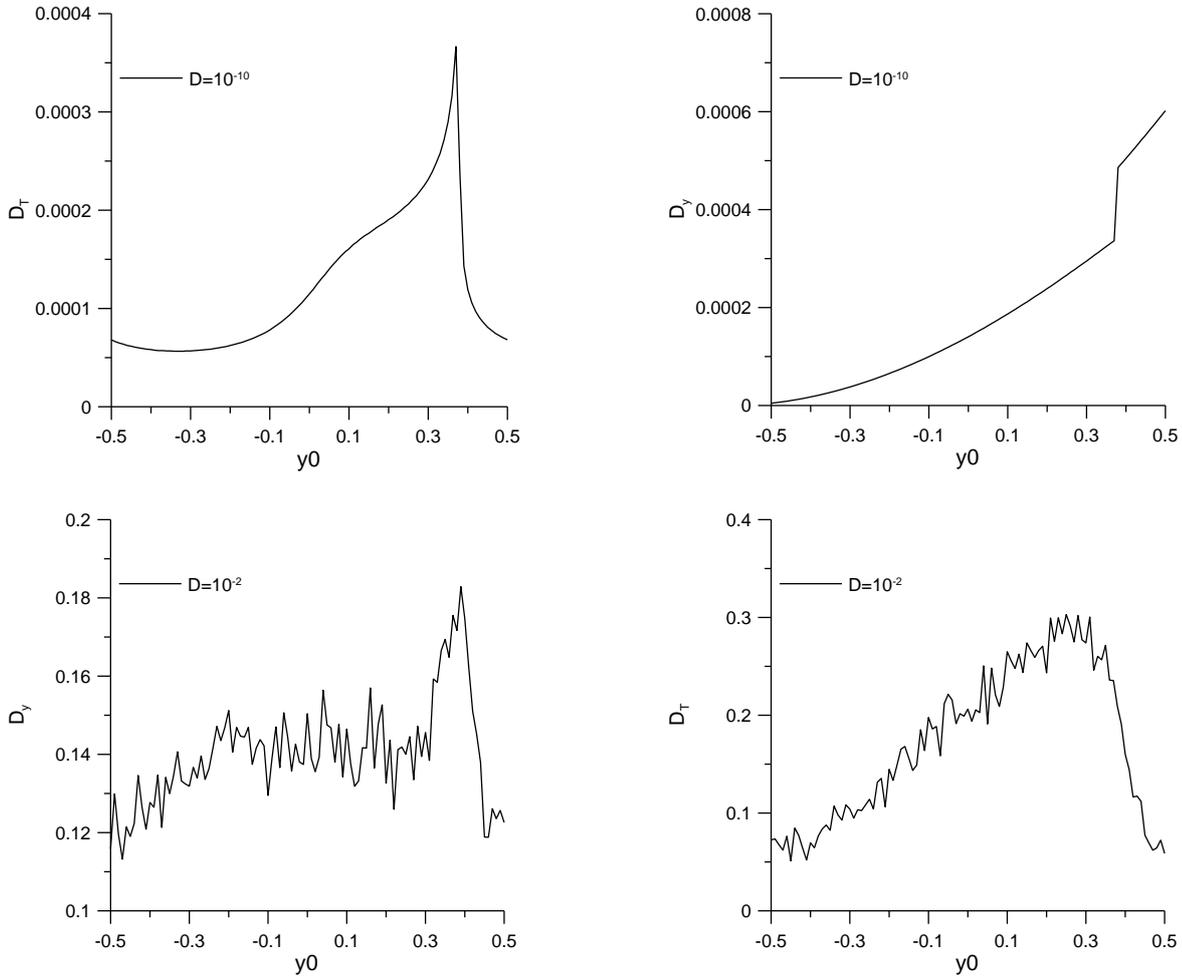

Fig 5 The dependence of the dispersions of time and coordinate on $y_0$ for different diffusivity of Gaussian noise

The behavior changes dramatically if one is consider non-Gaussian noise. Indeed, substituting $G^i, Q^i$ as the independent random variables with Cauchy distribution in Eq. (13), we repeated the same procedure described above. The obtained averaging functions $T(y_0)$ and $y_1 = H(y_0)$ are shown in Fig.7. It is seen that at small scale parameter $D$ the dependence $T(y_0)$ is the same as for Gaussian noise (see fig. 5,a,b). The increasing of scale parameter $D$ leads to a huge homogenization of dependences.

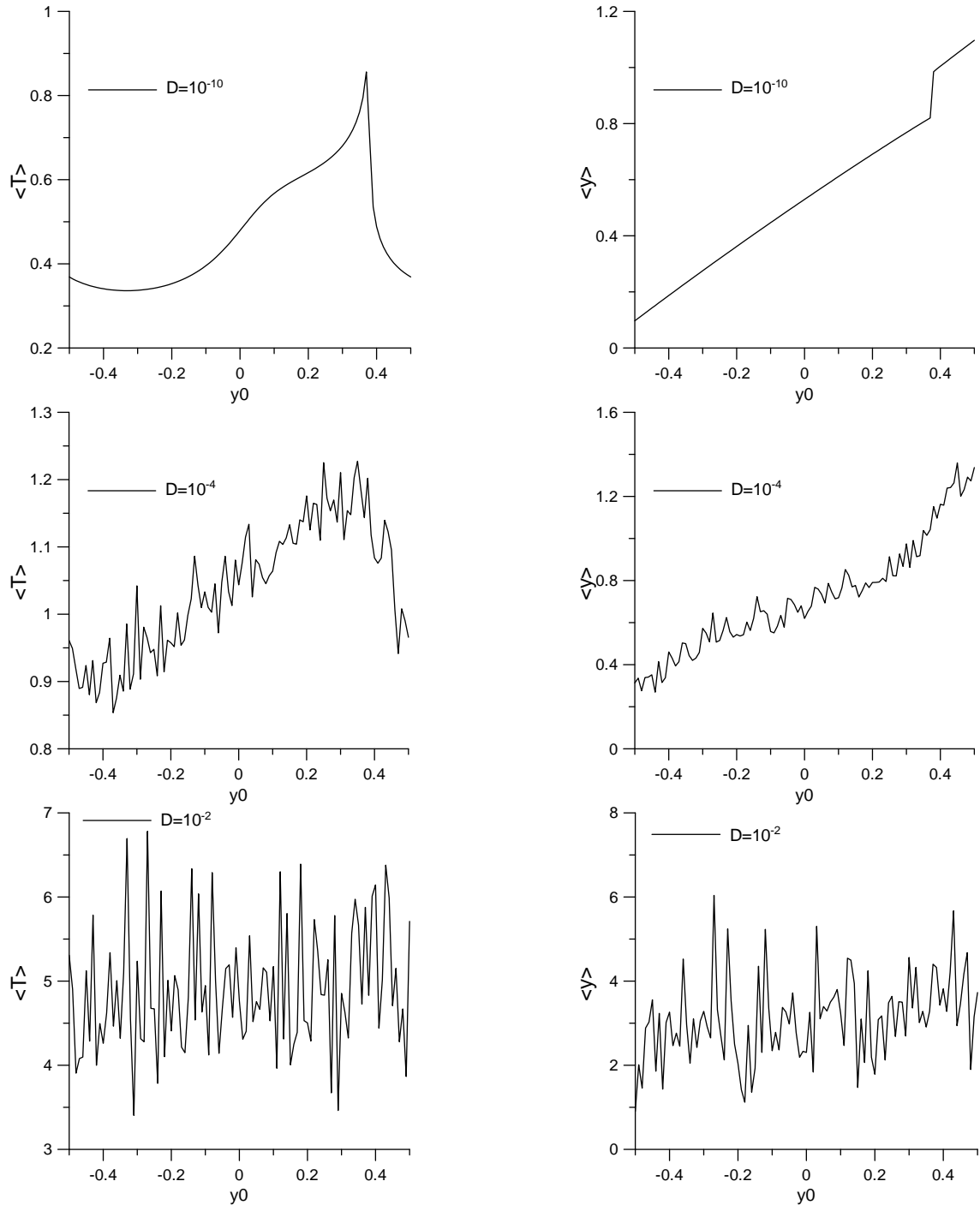

Fig 7 The dependence of the average time and coordinate on $y_0$ for different scale parameter of Cauchy noise

## 3. Conclusion

We have a made a generalization of the known [9] analytical solution of the system of equations describing two-dimensional viscous fluid flow in the presence of a force. The obtained family of solutions describes a two-dimensional viscous fluid flow periodic in both directions. The periodicity is imposed by the structure of the external force. It is shown that when the force amplitude reaches a certain critical value, a single closed vortex appears in each "cell". This value depends on the wavelength of the structure in both directions and the viscosity of the liquid. Thus, a periodic structure of asymmetric vortices is obtained. The transport of an ensemble of passive particles through the obtained structure is studied. The envelope of each vortex contains a stagnation point, i.e., the motion along it for a solitary passive particle in the absence of noise will be infinitely long. Since transport on an infinite periodic structure (lattice) is modeled, and the flow is set in such a way that resonant closure (periodicity) of trajectories is excluded (the ratio of flow rates in x- and y-directions is irrational), any trajectory of a solitary particle will pass infinitely close to the stagnation point.

Thus, it is shown that macroscopic transport of an ensemble of particles through such a lattice is possible only when internal noise is taken into account. The latter allows to "stir" the ensemble of particles, not letting the particles stay too long at the stagnation point. To investigate the characteristics of such transport, a model of "special flow" is constructed [8]. The model is a series of recurrence relations. Such a description is possible because for a periodic lattice all characteristics of a single particle at crossing the boundary of any unit cell are a function of the coordinate of the "entrance" of the particle into the previous cell. The recurrence relations are constructed by numerical solution of the Langevin equations in the presence of a random force, for an ensemble of passive particles during transport through a solitary cell.

**Funding Statement:** The authors acknowledge the financial support from RSF (Grant no. 23-12-00180)